# Predicting Economic Welfare with Images on Wealth

Jeonggil Song[1]

June 29, 2022

## Abstract

Using images containing information on wealth, this research investigates that pictures are capable of reliably predicting the economic prosperity of households. Without surveys on wealth-related information and human-made standard of wealth quality that the traditional wealth-based approach relied on, this novel approach makes use of only images posted on Dollar Street as input data on household wealth across 66 countries and predicts the consumption or income level of each household using the Convolutional Neural Network (CNN) method. The best result predicts the log of consumption level with root mean squared error of 0.66 and R-squared of 0.80 in CNN regression problem. In addition, this simple model also performs well in classifying extreme poverty with an accuracy of 0.87 and F-beta score of 0.86. Since the model shows a higher performance in the extreme poverty classification when I applied the different threshold of poverty lines to countries by their income group, it is suggested that the decision of the World Bank to define poverty lines differently by income group was valid.

[1] Sogang University, School of Economics, Baekbeom-ro 35, Mapo-gu, Seoul, South Korea (E-mail address: jgsong@sogang.ac.kr).

# 1. Introduction

Consumption or income is one of the most essential variables in economics that represents the socio-economic welfare of households. These data are also important in terms of calculating the poverty headcount ratio and the poverty targeting. At the same time, however, this kind of data is never easily accessible, especially in developing countries. Also, as Lindgren pointed out in his article "Detailed income calculations for Dollar Street", household income inappropriately captures the economic situations of people living in poor countries because of unstable income, in-kind payment, black economy, and the absence of labor contracts. These problems cast doubt on the national statistics or survey on income in developing countries.

Alternatively, since the late 1990s, many researchers used wealth indices for measuring socio-economic status of households in low and middle income countries and studying variation in health, mortality, poverty, education, work and other outcomes. It was found out by several researchers that wealth indices are suitable indicators of long-term socio-economic prosperity of households and they perform as well or better than expenditure data in capturing variation in education (Filmer and Pritchett 1999, 2001). Easy calculation and intuitive interpretation of the wealth index were the keys to appeal to many economists and practitioners. In addition, the fact that this wealth-based approach was applicable with previous household surveys such as Demographic and Health Surveys (DHS) and UNICEF MICS surveys was the reason for the success of the wealth index. (Smits and Steendijk 2015)

Although these useful properties of the wealth index played an important role in making it widely applicable, the wealth index suffers from one great limitation by nature: they heavily depend on survey data. Because the wealth index is calculated based on information related to whether there is a specific asset in a household, the index can be derived only for a household where there is survey data on wealth. Furthermore, in a traditional wealth-based approach, surveys on wealth which are used for calculating the wealth index utilize human-made standards of wealth quality such as floor material, toilet facility, and water source, so this subjective classification of wealth quality may harm the reliability of the index.

Unlike the traditional wealth-based approach which uses surveys on wealth and human-made standard of wealth quality, this research investigates the capability of images to reliably predict the socio-economic status of households by using only images containing information on wealth such as toilets, roofs, and stoves as inputs of the deep learning model. Because the input data are images itself and the standards of wealth quality are not necessary when training the model, this novel approach avoids the issues of dependency on survey data and of subjectivity of quality-related standards. This research makes use of images posted on Dollar Street as input data on household wealth across 66 countries and predicts the consumption or income level of each household using the Convolutional Neural Networks (CNN) method.

The merged image which leverages all information of images of seven wealth categories yields the best result. The best result predicts the log of consumption level with root mean squared error of 0.66 and R-squared of 0.80 in CNN regression problem. Among seven categories used in the analysis, the results of stoves and bathrooms are the best when using images of each category as inputs of CNN. In addition, this simple model also performs well in classifying extreme poverty with an accuracy of 0.87 and F-beta score of 0.86. Since the model shows a higher performance in the extreme poverty classification when I applied the different threshold of poverty lines to countries by their income group, it is suggested that the decision of the World Bank to define poverty lines differently by income group was valid. The results also imply that by taking a wealth-based approach and presenting a decent predictive power of wealth-related images on consumption level, this result supports the potentials of the wealth index as a proxy for consumption expenditures.

## 2. Data

I used photos posted on Dollar Street as input data and information on monthly consumption and country of each household. Dollar Street is a web page posting photos and videos related to the wealth of each household with information on monthly consumption (or income), country and short description of the household. Dollar Street was invented by Anna Rosling Rönnlund at Gapminder. Gapminder is an independent Swedish foundation with no political, religious or economic affiliations. They fight misconceptions surrounding global development with a fact-based worldview based on reliable statistics. Anna explains: "People in other cultures are often portrayed as scary or exotic. … We want to show how people really live. It seemed natural to use photos as data so people can see for themselves what life looks like on different income levels."

The Dollar Street team features 430 families in 66 countries with 44581 photos and 7352 videos and the list is growing. In each home, the photographer spends a day taking pictures of up to 135 objects such as the toilet, toothbrushes, and the roof. In addition, the photographer asks each household member several questions to figure out her/his average income, family size, age, and the proportion of food expenditure to the entire budget as well as the profession and the size of own food production.

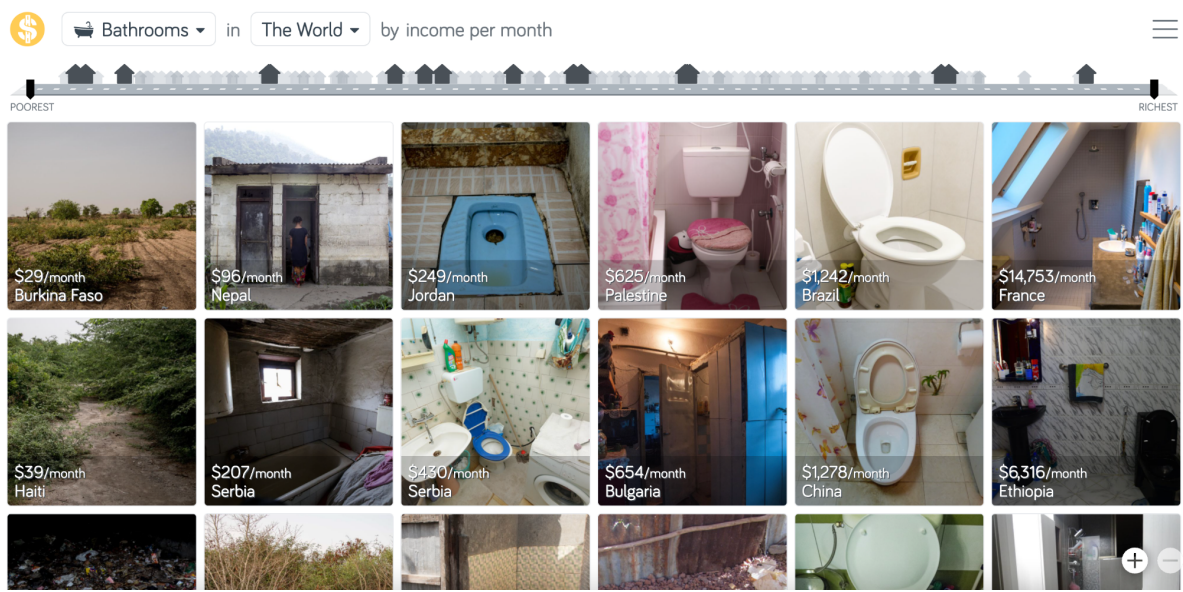

Figure 1. Dollar Street. https://www.gapminder.org/dollar-street?topic=bathrooms

What the Dollar Street team aimed to measure in principle was the living standard of each household during the last year the home was photographed. With this purpose in mind, they decided to measure consumption, not income, because consumption can be smoothed with savings or borrowing so that consumption better represents long-term living standards. Also, they divide total consumption with the number of so-called "adult equivalents". To take into account the fact that children require less spending than adults, and that there are economies of scale with bigger households, many surveys use equivalence scales. Using OECD-modified scale, they compute the number of "adult equivalents" by assigning a value of 1 to the household head, of 0.5 to each additional adult member and of 0.3 to each child (under the age of 14 years).

Even though their ambition is to display the consumption, over an extended period (a year), and express it as per adult equivalent, the availability of the information often hinders their ambition. However, they tried to recover the monthly consumption per adult equivalents by using the alternative and indirect information from a questionnaire. Other sources of indirect information include the family member's job, the average income level of specific professions in a specific country, reported assets, and data on minimum wages for each country. Since they report the income of households instead of the consumption level in some cases where they can't even guesstimate in spite of all indirect sources of information, I will not discriminate between both terms clearly in this paper. One thing to note in this point is that the Dollar Street team put great efforts into estimating consumption as precisely as they could.

## 3. Prediction Methods

Recently, the state-of-the-art machine learning techniques have become easier to use even for social scientists and these methods often appear in economic research in order to solve the problem of scarce data. For example, Huang et al. (2021) evaluated the impact of anti-poverty programs by leveraging satellite imagery and deep learning models. Also, Dell et al. (2021) proposed a deep learning library named LayoutParser which is used for extracting useful data from document images.

In this research, the deep learning model predicts the log of consumption in the regression problem and extreme poverty in the classification problem using the Dollar Street images related to the wealth of each household. The whole procedure of the experiment consists of three main steps. First, web scraping, which is a data scraping used for extracting useful data from websites, needs to be done to prepare input images and information on the consumption level and country for the analysis. Second, I extract the information on consumption levels and country from the name of image files which were downloaded from web scraping. Also, during the pre-processing task, I label each household into extreme poverty based on monthly consumption and country information and merge 7 categories of images to leverage all information of image inputs. Finally, the CNN model is trained to predict the log consumption level and extreme poverty label of each household using input images and y labels from the previous steps.

- Web scraping and downloading images

Even though downloading images from Dollar Street is possible, Dollar Street only provides non-square pictures with too high resolution (e.g. 2799*1865 pixels, 493KB) and, more importantly, it is not available to download all images at once. To alleviate computational burden and reduce manual labor, it is necessary to perform web scraping to download images. Web scraping is a data scraping process that allows us to efficiently collect relevant data and provides us with the tools to do the job automatically. Furthermore, when downloading images using web scraping, it is also possible to save the name of images as the consumption value and country name of each image which are used as family ID and y-label.

Among 198 categories, only seven categories of images on wealth are used as input data, which are bathrooms, bedrooms, living rooms, places for dinner, roofs, showers, and stoves, based on three criteria: (1) The quality improvement of wealth is apparent as the level of income increases, (2) a type of wealth is so essential for living that every household has it, (3) the image is available in almost all of family in Dollar Street.

- Pre-processing

When the web scraping step is finished, the outputs are images of seven categories whose file name is its family ID. Based on the monthly consumption of each family, I generate an extreme poverty label which will be used as a y-label in a classification problem. If the monthly consumption of a household does not exceed 57 dollars ($1.9 times 30 days), the value of extreme poverty is one. If not, the value is zero. In the experiment to apply different poverty lines by each country's income group classified by the World Bank, different labels are used: the poverty lines for the low-income countries (LICs), the lower middle-income countries (LMICs), the upper middle-income countries (UMICs) and the high-income countries (HICs) are $1.9 / a day, $3.2 / a day, $5.5 / a day and $21.7 / a day, respectively.

In addition, merging images is necessary to test the predictive power of merged inputs. Merged images consist of all images of 7 categories. If a specific wealth image of a family is not available, then the designated location of that wealth is substituted by a white image. Merged images not only leverage all information of seven wealth categories, but also have the meaning of unifying the samples. Since each category has a different number of samples (images), the outcomes of the prediction of each category are not comparable. However, in the prediction of merged inputs, the result can be considered as a baseline since it utilizes all samples. As a result, the total dataset consists of 426 families, but 16 families which consume over 5000 dollars a month are dropped since they are assumed to be outliers. Therefore, the final dataset used in this analysis has 410 families.

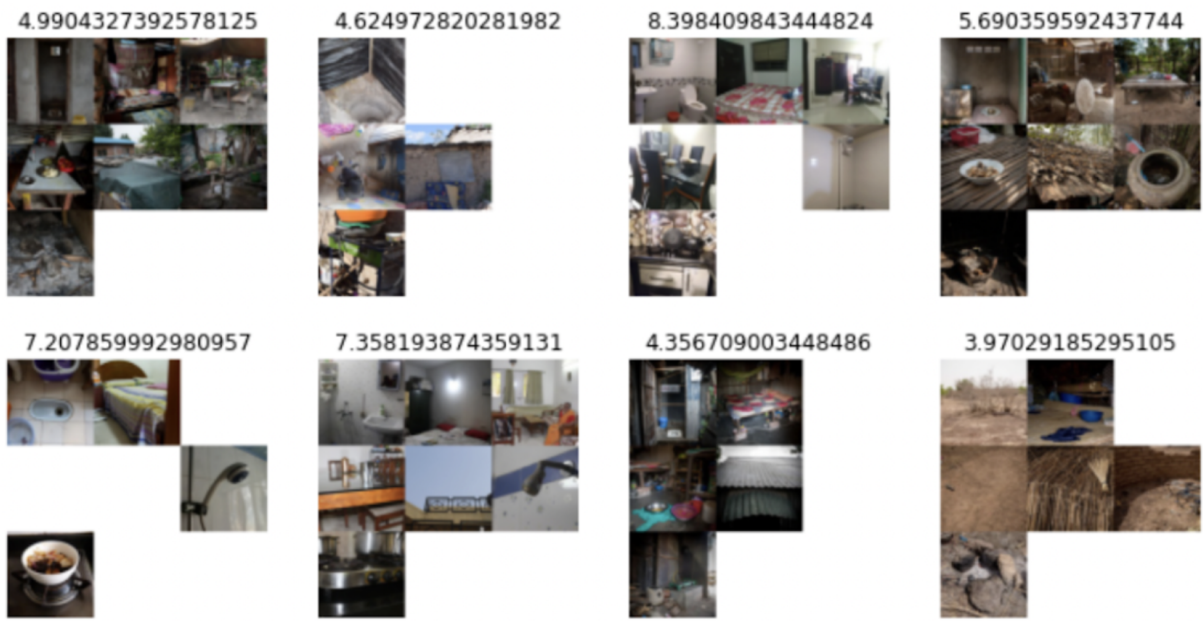

Figure 2. Samples of merged inputs of CNN. The number above the image is the ground truth of the log of monthly consumption. The order of images in merged input is bathrooms, bedrooms, living rooms, places for dinner, roofs, showers, stoves (left to right, then top to bottom)

- Training the models

I split the dataset into two parts for every experiment: 80% belongs to the train set and 20% belongs to the validation set. The main method which is used in this research to analyze images is the Convolutional Neural Networks (CNNs). Among many techniques, CNN method is one of the most popular deep learning methods in computer science. CNNs are a type of artificial neural networks that are specialized in and widely used for analyzing visual imagery. Using input images and ground-truth labels, the model is trained through multiple layers and extracts features that are important in each label. After the model learns about rules that are crucial in the given ground-truth labels, it predicts labels for unseen images. This CNN method can be easily applied to the regression problem for predicting continuous variables and classification problem for predicting categorical variables.

In this research, the CNN model is used to solve two problems. The first is the regression problem to predict the log of monthly consumption and the second is the classification problem to predict the extreme poverty label.

## 4. Results

In this section, I show the results of two problems: the CNN regression problem to predict the log of monthly consumption and the CNN classification problem to predict extreme poverty.

4.1 Regression

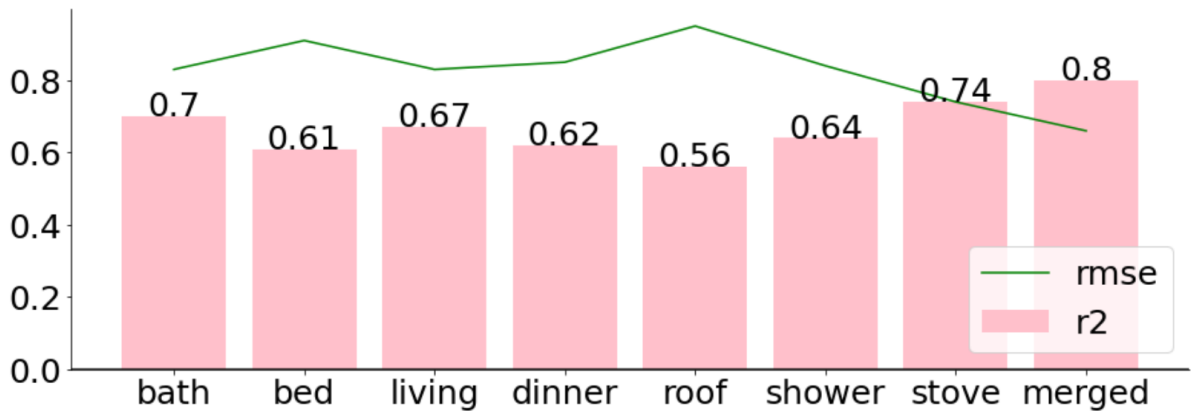

Figure 3. The results of CNN regression of each category. Categories are bathrooms, bedrooms, living rooms, places for dinner, roofs, showers, stoves, and merged input in order. "rmse" is the root of mean squared error and "r2" is R- squared value. The sample size of each category is 365, 382, 285, 369, 316, 344, 391, 410 in order from bathrooms to merged inputs.

I regress the predicted consumption level on the actual value and report R-squared value and the root of mean squared error (RMSE). Figure 3 shows the predictive power of each wealth category. I find that all categories of wealth items are strongly predictive of the log of consumption level. Especially, when using single input of category as inputs, images of stoves and bathrooms yield the best performance which are 0.7 of R-squared value and 0.83 of RMSE for bathrooms and 0.74 of R-squared value and 0.74 of RMSE for stoves.

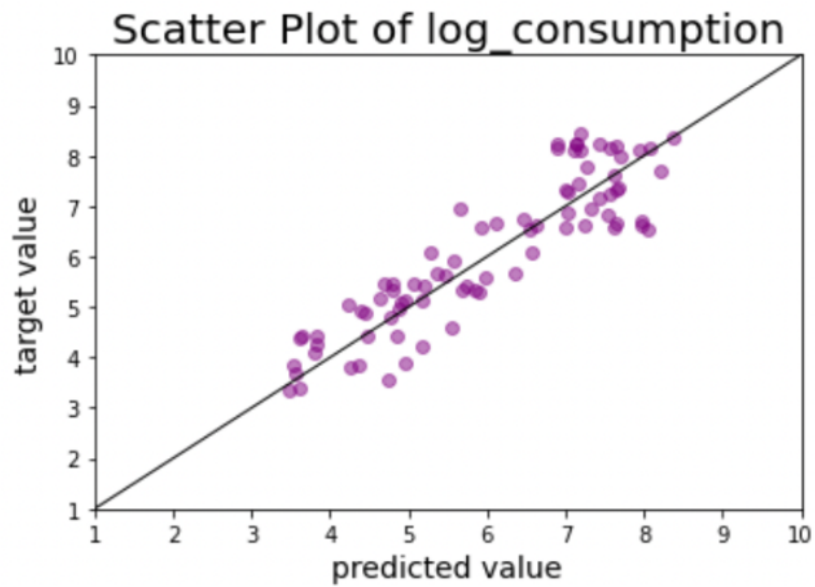

Figure 4. The scatter plot of merged input which yields the highest performance.

Interestingly, the merged inputs show the highest predictive power among all categories. The R-squared value is 0.8 and the RMSE is 0.66. Considering that RMSE represents the degree to which the result of predictions yields the error on average, 0.66 of RMSE seems pretty low. Figure 4 shows the result of predictions of merged inputs. The horizontal axis represents the predicted value and the vertical axis represents the target value, i.e., the ground-truth y-label. The diagonal solid line is the 45 degree line, so the prediction results are more reliable as the scatter plots are closely distributed around the 45 degree line.

4.2 Classification

- All countries (balanced) (n=225+225, beta=0.8)

| epoch | train_loss | valid_loss | accuracy | precision_score | recall_score | fbeta_score | time |
|---|---|---|---|---|---|---|---|
| 6 | 0.161974 | 0.670611 | 0.833333 | 0.795455 | 0.853659 | 0.817198 | 01:01 |
| 7 | 0.137514 | 0.707058 | 0.833333 | 0.795455 | 0.853659 | 0.817198 | 01:00 |
| 8 | 0.116265 | 0.766391 | 0.822222 | 0.777778 | 0.853659 | 0.805727 | 01:01 |
| 9 | 0.099293 | 0.791012 | 0.844444 | 0.800000 | 0.878049 | 0.828748 | 01:01 |
| 10 | 0.085429 | 0.807984 | 0.844444 | 0.800000 | 0.878049 | 0.828748 | 01:02 |
| 11 | 0.073886 | 0.826271 | 0.822222 | 0.765957 | 0.878049 | 0.806117 | 01:01 |
| 12 | 0.066343 | 0.868587 | 0.833333 | 0.770833 | 0.902439 | 0.817349 | 01:01 |

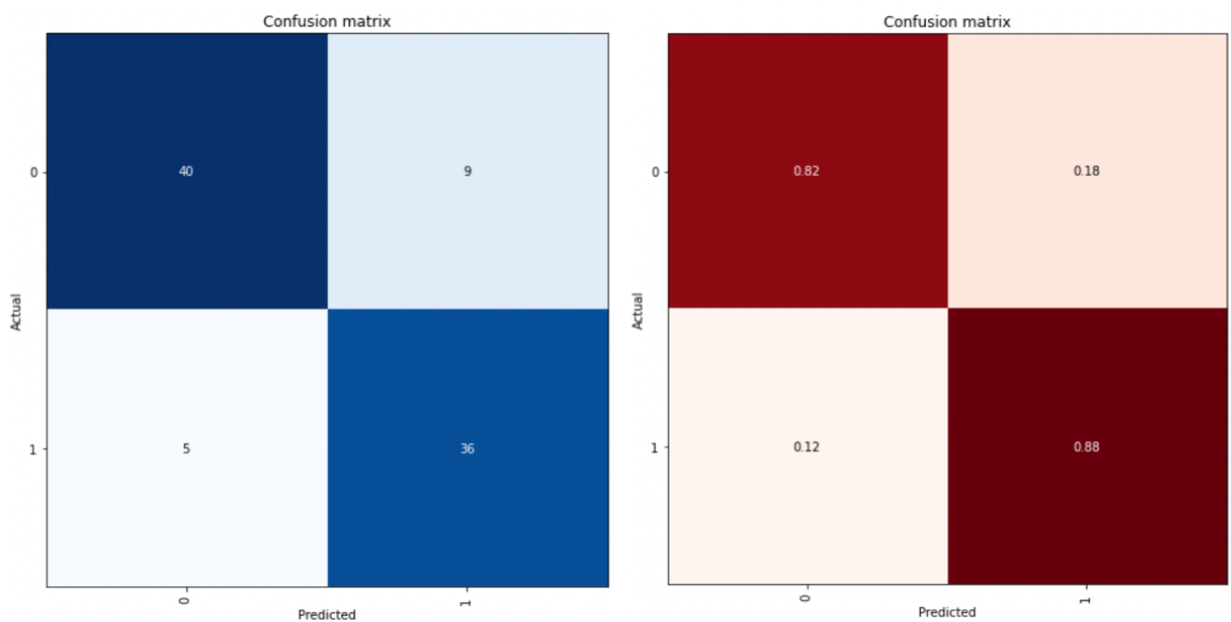

Figure 5. The result of extreme poverty classification with the same poverty line ($1.9 / a day). The label 0 represents the non-extreme poverty, and the label 1 represents the extreme poverty.

If the model is trained with an imbalanced dataset, the model may give biased predictions. Since there are only 225 images which belong to extreme poverty among 2562 images, I randomly choose 225 images among images which have non-extreme poverty labels in order to handle imbalance data issues. The accuracy of the model is 0.84, which means the model gives true predictions with probability of 84% among 90 images of the validation set. The colored figures are called confusion matrices. They show the results of predictions visually. As shown in the blue matrix, the model yields 40 true negatives out of 49 images of label 0, and 36 true positives out of 41 images of label 1. The red matrix is the normalized version of the blue matrix with the same result.

- By income group (balanced) (n=433+433, beta=0.8)

| epoch | train_loss | valid_loss | accuracy | precision_score | recall_score | fbeta_score | time |
|---|---|---|---|---|---|---|---|
| 12 | 0.034408 | 0.804400 | 0.826590 | 0.781250 | 0.892857 | 0.821314 | 01:55 |
| 13 | 0.028300 | 0.833261 | 0.849711 | 0.808511 | 0.904762 | 0.843530 | 01:56 |
| 14 | 0.023221 | 0.978902 | 0.832370 | 0.789474 | 0.892857 | 0.826835 | 01:55 |
| 15 | 0.019396 | 0.944283 | 0.867052 | 0.814433 | 0.940476 | 0.859379 | 01:57 |
| 16 | 0.015825 | 0.913227 | 0.861272 | 0.812500 | 0.928571 | 0.854167 | 01:56 |
| 17 | 0.013695 | 0.886863 | 0.855491 | 0.817204 | 0.904762 | 0.849278 | 01:57 |
| 18 | 0.011523 | 0.840207 | 0.838150 | 0.804348 | 0.880952 | 0.832602 | 01:57 |

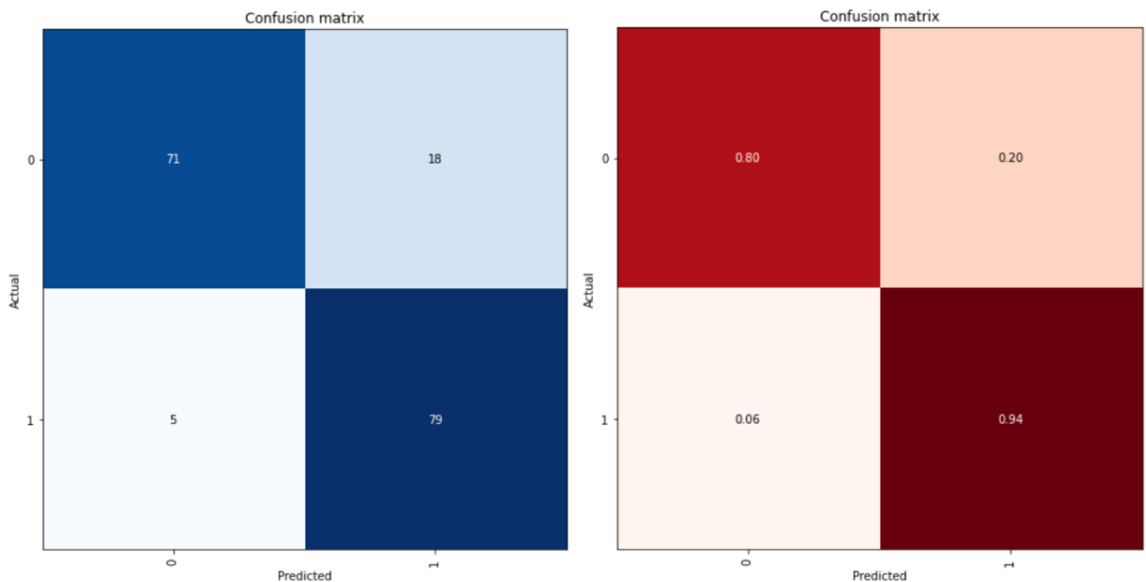

Figure 6. The result of extreme poverty classification with different poverty lines by income group. The poverty lines for the low-income countries (LICs), the lower middle-income countries (LMICs), the upper middle-income countries (UMICs) and the high-income countries (HICs) are $1.9 / a day, $3.2 / a day, $5.5 / a day and $21.7 / a day, respectively. What consists of each income group is described here.

Figure 6 shows the result of predictions when the different poverty lines by income group are applied to each country. In this experiment, the number of images of extreme poverty is 433, so I randomly choose 433 images among images which have non-extreme poverty labels for the same reason stated above. The accuracy of the model is 0.87, which means the model gives true predictions with probability of 87% among 173 images of the validation set. As shown in the blue confusion matrix, the model yields 71 true negatives out of 89 images of label 0, and 79 true positives out of 84 images of label 1. Again, the red matrix is the normalized version of the blue matrix with the same result.

Compared to the results of the single poverty line ($1.9 / a day), the performance of the model has improved in terms of all scores when the different poverty thresholds are applied

to each income group. However, the results are not directly comparable since they use the different samples, that is, the different train set and the different validation set. Nevertheless, through the repeated experiments and the consistent results (CHECK!), we may be able to interpret this result as an implication that the decision of the World Bank to define poverty lines differently by income group was valid.

## 5. Conclusions

In this paper, I show that images containing information on wealth can be used to predict the socio-economic status of each household, specifically the consumption level. This result implies that practitioners and researchers can leverage images to get predictions on the consumption data of each household with purposes of the poverty targeting or poverty assessment. For example, the households with predictions of extreme poverty can be the first candidates for the cash transfer. This method of measuring poverty may be more reliable than the survey considering that it is pointed out that the responses of households get changed depending on the contents of the questionnaire and many cast doubt on the reliability of the survey in this context because of the false report of households to get assistance.

Also, this result can be applied to give predictions of consumption data for broader regions without survey data by using google street view as inputs. Once checking the predictive power of images which are taken outside the house, the validity of this application would be guaranteed. Considering the scarcity of data in developing worlds, these applications can have meaningful contributions for pragmatic purposes such as poverty mapping.

The limitation of this research is that the results of CNN models of each category are not comparable since they all use different samples. Also, in a classification problem, it is not appropriate to conclude that the way of the World Bank to define poverty lines differently is valid since the label for some images get changed so that it is impossible to use the same samples in this case. Thus, in order to choose the wealth category which yields the best performance, it will be the next step of the research to make results comparable by using the same samples in the model training and validation.